\documentclass[conference]{IEEEtran}
\IEEEoverridecommandlockouts

\usepackage{cite}
\usepackage{amsmath,amssymb,amsfonts}
\usepackage{algorithm}
\usepackage{algpseudocode}
\usepackage{graphicx}
\usepackage{textcomp}
\usepackage{multirow}
\usepackage{csquotes}
\usepackage{hyperref}
\usepackage{listings}
\usepackage{lettrine}
\usepackage{array}
\usepackage{stfloats}
\usepackage{url}
\usepackage{verbatim}
\usepackage{academicons}
\usepackage{upgreek}
\usepackage{siunitx}
\usepackage{adjustbox}
\usepackage{rotating}
\usepackage[inline]{enumitem}
\usepackage{relsize}
\usepackage{longtable}
\usepackage{xcolor}
\usepackage{threeparttable}
\usepackage[caption=false]{subfig}
\usepackage{tabu}
\usepackage{color, colortbl}
\usepackage{booktabs}
 
\usepackage{pifont}
\usepackage{tabularx}

\begin{document}

\title{A Review of Memory Wall for Neuromorphic Computing}

\makeatletter
\newcommand{\linebreakand}{%
  \end{@IEEEauthorhalign}
  \hfill\mbox{}\par
  \mbox{}\hfill\begin{@IEEEauthorhalign}
}
\makeatother

\author{%

    \IEEEauthorblockN{Dexter Le}
    \IEEEauthorblockA{\textit{Drexel University} \\
    Philadelphia, USA \\
    dql27@drexel.edu \\
    }

    \and

    \IEEEauthorblockN{Baran Arig}
    \IEEEauthorblockA{\textit{Drexel University} \\
    Philadelphia, USA \\
    ba646@drexel.edu \\
    }

    \and

    \and
    \IEEEauthorblockN{Murat Isik}
    \IEEEauthorblockA{\textit{Stanford University} \\
    Stanford, USA \\
    misik@stanford.edu}
    
    \linebreakand
    
    \IEEEauthorblockN{I. Can Dikmen}
    \IEEEauthorblockA{\textit{Yildiz Technical University} \\
    Istanbul, Turkey \\
    can.dikmen@yildiz.edu}
        \and
    \IEEEauthorblockN{Teoman Karadag}
    \IEEEauthorblockA{\textit{Temsa Research \& Development Center} \\
    Adana, Turkey \\
    teoman.karadag@temsa.com}

 }   
\maketitle


\maketitle

\vspace{-10pt}

\begin{abstract}
This paper reviews memory technologies used in Field-Programmable Gate Arrays (FPGAs) for neuromorphic computing, a brain-inspired approach transforming artificial intelligence with improved efficiency and performance. It focuses on the essential role of memory in FPGA-based neuromorphic systems, evaluating memory types such as Static Random-Access Memory (SRAM), Dynamic Random-Access Memory (DRAM), High-Bandwidth Memory (HBM), and emerging non-volatile memories like Resistive RAM (ReRAM) and Phase-Change Memory (PCM). These technologies are analyzed based on latency, bandwidth, power consumption, density, and scalability to assess their suitability for storing and processing neural network models and synaptic weights. The review provides a comparative analysis of their strengths and limitations, supported by case studies illustrating real-world implementations and performance outcomes. This review offers insights to guide researchers and practitioners in selecting and optimizing memory technologies, enhancing the performance and energy efficiency of FPGA-based neuromorphic platforms, and advancing applications in artificial intelligence.
\end{abstract}

\section{Introduction}

In the era of big data, artificial neural networks (ANNs) have emerged as a cornerstone in the field of artificial intelligence (AI), demonstrating remarkable success across diverse applications. Despite their claim of being inspired by biological neural networks, ANNs, particularly those employing backpropagation algorithms, diverge significantly in operation from the human brain \cite{isik2024advancing}. As state-of-the-art neural network models expand to tackle increasingly complex tasks, the issue of power efficiency becomes paramount. In this context, recent advancements in brain-inspired computing, notably neuromorphic chips and Spiking Neural Networks (SNNs) \cite{Maass1997}, represent a paradigm shift in AI hardware and software. These technologies, mirroring the principles of the human brain, have shown superior efficiency and effectiveness compared to conventional neural networks in terms of both accuracy and learning capabilities. However, the development of neuromorphic computing hardware remains a niche, with limited manufacturers, rendering access to such technology challenging. An alternative to the specialized and scarce Application-Specific Integrated Circuits (ASICs) used in neuromorphic computing is the use of Field-Programmable Gate Arrays (FPGAs). FPGAs, known for their asynchronous nature, offer a versatile and cost-effective platform for implementing and testing neuromorphic algorithms. This reconfigurable hardware presents a more accessible option for a broader range of users, compared to neuromorphic ASICs.

A critical aspect of implementing neuromorphic solutions on FPGAs involves the storage of model data (weights, connections) in memory for integration into crossbar arrays. This paper aims to review various memory technologies, analyzing their respective advantages and disadvantages. Popular memory choices include Static Random Access Memory (SRAM), Dynamic Random Access Memory (DRAM), and, more recently, High-Bandwidth Memory (HBM) \cite{hbm_neuromorphic}. The selection of memory type—volatile for storing cross-bar weights and non-volatile for the complete model and weights—depends on the specific characteristics of each memory technology.

Notably, neuromorphic devices such as Loihi \cite{loihi} and TrueNorth \cite{truenorth} predominantly utilize SRAM for weight storage. SRAM's low-latency characteristic is advantageous, yet the limited bandwidth capabilities of FPGAs necessitate exploring alternative memory solutions for achieving higher density. High-bandwidth memory emerges as a potential candidate, allowing for the stacking of multiple DRAMs, which can be accessed in parallel—a feature well-suited to the parallel architecture of FPGAs. Furthermore, HBM's high-bandwidth capacity enables a hierarchical approach to loading model weights onto crossbars \cite{coreset}.

Memory design patterns in SNNs can be implemented to reduce DRAM accesses through near-memory processing. Other patterns in improving storage density may also be accomplished with SRAM through Read-Only Memory (ROM) cache integrations, which also contribute to mitigating the memory wall. The memory-wall problem is a phenomenon where the processor's speed outpaces the transfer of data through memory. In traditional computing, typical mitigations for the memory-wall problem involve utilizing an L2 cache and avoiding memory fetches when possible. Despite this mitigation, present bottlenecks are potentially within the chip memory \cite{gholami2024aimemorywall}. Neuromorphic computing aims to solve this due to the inherent parallelism of operations. Given neuromorphic computing, FPGAs also present a potential for adoption for neuromorphic systems where environmental constraints such as power consumption and speed must be held paramount. Multiple blocks of RAM are incorporated which enable synaptic weights to be accessed concurrently. As a result, parallelized operations on the FPGA greatly improve energy efficiency and speed. 

\begin{figure}
    \graphicspath{ {D:\Stack} }
    \center \includegraphics[width=0.50\textwidth]{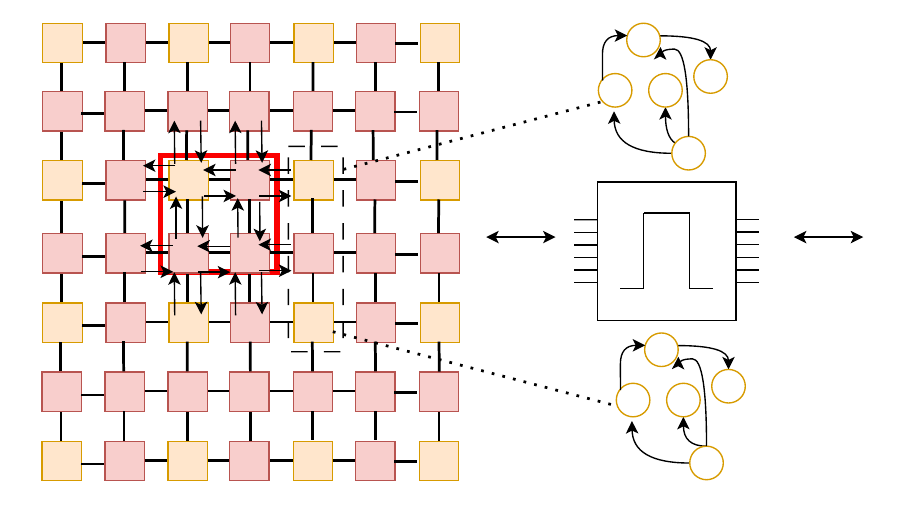}
    \caption{Hierarchical Dataflow in Neural Network Accelerator with Layered Memory and Processing Units.  }
    \label{test4}
\end{figure}

Fig. \ref{test4} shows a hierarchical dataflow architecture used in a neural network accelerator, designed to optimize the movement and processing of data across multiple layers of memory and computational units. The structure emphasizes the efficient handling of large neural network models by leveraging a combination of high-speed memory and processing elements. It is ideal for deep learning applications where managing large datasets and computational workloads efficiently is essential for high performance and scalability.

\section{Technologies}

The ability to parameterizable, accurate, and balanced performance in comparison to the software platform is the primary goal for building a hardware implementation in embedded systems. According to the findings, QR decomposition and modified Gram–Schmidt methods may deal with matrix inversion approaches, which maximize resource utilization and stability in embedded systems. Furthermore, the training module, ANN module, RAM memory for storing, and data flow will be the expert system's four primary hardware structure blocks.

\begin{figure}
    \graphicspath{ {D:\Stack} }
    \center \includegraphics[width=0.50\textwidth]{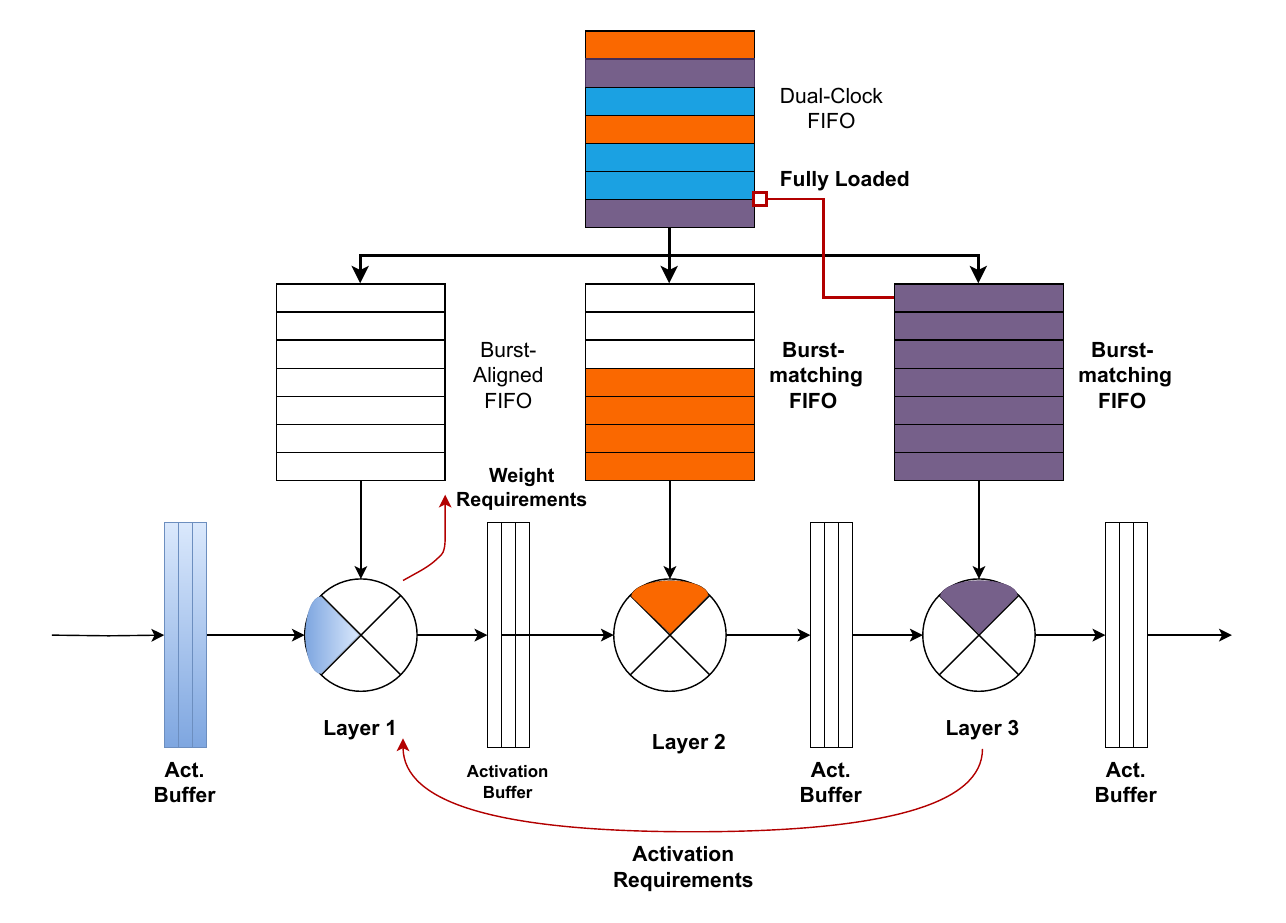}
    \caption{Dataflow Architecture for Neural Network Processing with FIFO Buffers and Activation Management.  }
    \label{test1}
\end{figure}

Fig. \ref{test1} illustrates a layered dataflow architecture optimized for neural network processing, with a focus on managing activations and weights. Multiple layers interact to handle data transmission and synchronization across processing stages. This design ensures efficient data management, enabling synchronized data transfer, storage, and processing to optimize memory usage and computational throughput. It is especially relevant for FPGA-based or hardware-accelerated networks, where efficient management of latency, bandwidth, and resources is essential for system performance.

\subsection{SRAM}

SRAM cells contain configuration data and program the routing connection. In SRAM-based FPGA, SRAM Cell output can regulate the functionality bits of logic blocks and their interconnections. \cite{bobda2007introduction}. SRAM can be utilized to configure and link logic blocks.  The main advantage of SRAM-based modules is that FPGAs may be reprogrammed or customized several times. The SRAM cells' values can be altered to create a new connection or function. SRAM cells occupy a significant chip area, adding to the technology's complexity. Because the device is volatile, data saved in static memory is lost if the power goes out \cite{farooq2012tree}. Non-volatile memory or external sources are required at every power-up to maintain the configuration data and save it in the FPGA device.

\subsection{BRAM and URAM}

Authors \cite{li2018general} provides a high-performance universal framework for speeding up Swarm intelligence algorithms (SIA). SIA reschedules the dataflow of SIAs to reduce data transfer between memory tiers, improving overall performance. SIA on FPGAs delivers higher throughput due to the distributed memory architecture and the tailored deep variable pipeline of FPAGs. When compared to the Intel Core i7-6700 CPU/ NVIDIA GTX1080 GPU, the proposed design is up to 123 times and not less than 1.45 times quicker in terms of optimization time on Xilinx XCKU040. The proposed architecture supports \cite{valavi201964} Analog/binary IA/weight first layer (FL) and binary/binary IA/weight hidden layers (HLs), with batch normalization and input-output (IO) (buffering) circuitry to permit cascading if required, for implementing alternative DNN layers. The architecture is organized as 64 in-memory computational neuron tiles, each of which may handle up to 512. A Python interface is provided by a host PC for transmitting and receiving instructions to and from an FPGA.

\subsection{DRAM}
The cost of monolithic DRAM per bit is approximately 100 times less than the cost of the densest logic-based memory. Traditional architectures optimize costs by splitting memory between cheap DRAM and expensive logic \cite{davies2021advancing}. However, external DRAM introduces bottlenecks, such as access latency and limited bandwidth, which impact accelerator performance \cite{wei2021memory}, \cite{bojnordi2013programmable}. While DRAM capacity and bandwidth have improved over generations, access latency has seen minimal progress. Accessing main memory can consume many clock cycles, resulting in idle times of up to 70\% on FPGA accelerators \cite{hussain2014advanced}, \cite{hussain2012ppmc}. To reduce latency, researchers propose batching memory requests based on spatial proximity and reordering them efficiently using FPGA parallelism, which increases DRAM row buffer hits \cite{ausavarungnirun2012staged}. The NNAMC memory controller for neural network accelerators, implemented on a Xilinx VC707 FPGA, dynamically maps memory access streams to reduce latency. This design increased row buffer hits by 13.68\% and reduced system access latency by 26.3\%, with improvements reaching 37.68\% in some cases \cite{wei2021memory}. Other studies focused on optimizing CNN computations and memory access patterns. Using a roofline model, researchers explored configurations and built DRAM implementations on the Xilinx VC707 board to optimize cross-layer design \cite{zhang2015efficient}. To further lower costs, researchers suggest future integration of emerging memory technologies such as ReRAM, MRAM, and PCM, which could store multiple bits per device while maintaining density and cost efficiency. In DNA sequencing, FPGA-based accelerators improved short-read mapping by leveraging DRAM for reference genome storage. A single human genome requires 22 GB of index and reference material, distributed across eight M-503s with 4 GB of DDR3 DRAM each \cite{olson2012hardware}. This division ensures efficient storage and access to pointer tables, reference data, and candidate alignment locations. Researchers also proposed a modular, programmable memory controller for DRAM on FPGA accelerators. This controller allows users to optimize external memory access while reusing hardware resources \cite{wijeratne2021programmable}. Studies like FP-DNN allocate DRAM buffers for implementation but struggle to utilize on-chip memory efficiently \cite{guan2017fp}. Similarly, Argus and DNNBuilder provide complete multi-layer accelerator architectures, though data overflow from BRAMs often requires off-chip memory storage \cite{shen2019argus}, \cite{zhang2018dnnbuilder}. The Gemmini project offers a modular DNN accelerator template with a complete hardware-software stack, including SoC integration \cite{genc2021gemmini}. It uses a systolic array for efficient data transfer between DRAM, cache, and scratchpad memory. Gemmini accelerators outperform commercial DNN accelerators and CPUs by orders of magnitude, demonstrating the potential of optimized DRAM-based designs.

\subsection{HBM}
The integration of high-bandwidth memory (HBM) on the same chip as an FPGA allows us to build our accelerator circuitry considerably closer to the memory while providing an order of magnitude more bandwidth than standard DDR4-based FPGA boards. As a result, contemporary FPGAs take a more datacentric approach to computation. Researchers \cite{singh2021fpga} used an FPGA with HBM to improve the pre-alignment filtering stage of genomic analysis as well as representative kernels from a weather prediction model. Tests show significant speedups and energy savings when compared to a high-end IBM POWER9 machine and a standard FPGA board with DDR4 memory. FPGA-based near-memory computing reduces the data transportation bottleneck in current data-intensive applications.

\begin{figure}
    \graphicspath{ {D:\Stack} }
    \center \includegraphics[width=0.50\textwidth]{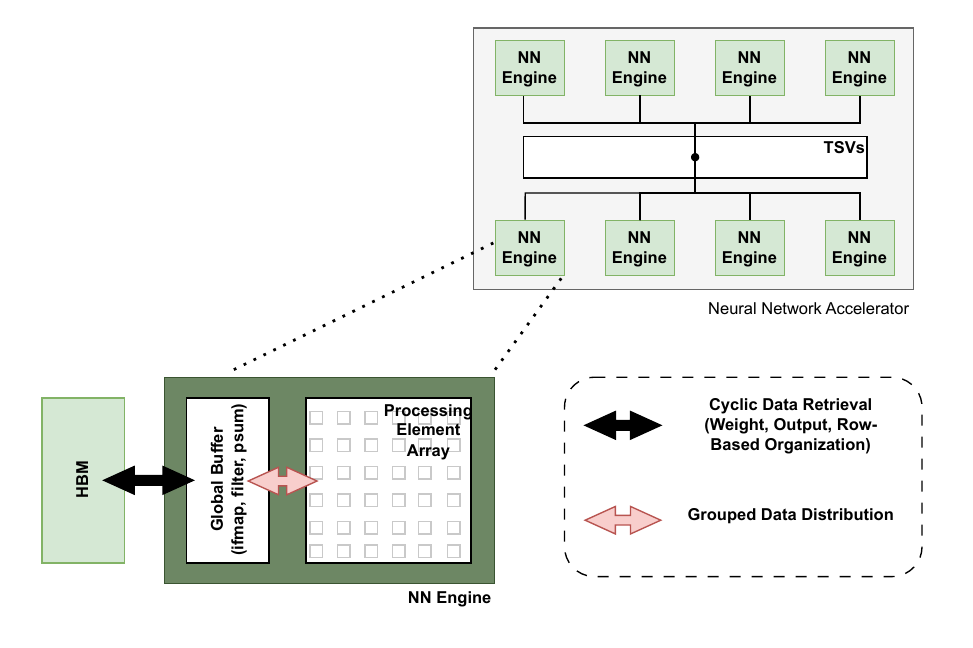}
    \caption{Neural Network Accelerator Architecture with Processing Elements and High-Bandwidth Memory Integration.  }
    \label{test2}
\end{figure}

Fig. \ref{test2} presents the architecture of a neural network accelerator optimized for efficient data processing and memory utilization in hardware-accelerated machine learning. Key components work together to enhance neural network operations. The architecture includes multiple Neural Network (NN) Engines and a Processing Element (PE) Array for matrix multiplications and convolutions. Through-Silicon Vias (TSVs) facilitate high-speed communication between NN engines and the High-Bandwidth Memory (HBM) stack, offering fast, low-latency access to large datasets and weights. A Global Buffer stores intermediate data, such as feature maps and filters, improving data reuse and minimizing memory accesses. Cyclic Data Retrieval fetches weights and outputs in a round-robin manner, while Grouped Data Distribution organizes data transfer row-wise, reducing latency and bandwidth demands. This design maximizes parallelism and memory efficiency, making it well-suited for deep learning tasks.

\subsection{Comparison of Memory Types}

\subsubsection{SRAM vs. DRAM}

SRAM offers low latency and high-speed access, making it ideal for on-chip storage of critical data such as crossbar weights and neuron states~\cite{tostado2019performance, pedroni2019memory}. However, SRAM has lower density and higher static power consumption due to its six-transistor cell design. This limits its scalability for large-scale neuromorphic networks. DRAM, on the other hand, provides higher density at a lower cost, suitable for off-chip storage of large synaptic weight matrices~\cite{alam2024achieving}. DRAM suffers from higher latency and requires periodic refresh cycles to maintain data integrity, which can introduce performance overheads. The destructive read operations in DRAM also necessitate write-back mechanisms, complicating memory management.

\subsubsection{Flash Memory}

Flash Memory is a non-volatile memory offering high density and data persistence. However, it has slower write speeds and limited write endurance, making it less suitable for applications requiring frequent weight updates, such as online learning in neuromorphic systems. Flash memory is more appropriate for storing static configuration data or infrequently updated parameters.

\subsubsection{Emerging Memory Technologies}

Phase-Change Memory (PCM), Resistive RAM (ReRAM), and Magnetic RAM (MRAM) are emerging non-volatile memory technologies that combine high density with faster access times. They offer potential advantages such as low power consumption, high endurance, and the ability to perform in-memory computations.

\begin{itemize} \item \textbf{PCM}: Utilizes the reversible phase change between amorphous and crystalline states to store data. PCM offers good scalability and moderate access speeds but suffers from high write energy and limited endurance~\cite{burr2010inner}. \item \textbf{ReRAM}: Operates by changing the resistance across a dielectric solid-state material. ReRAM boasts fast switching speeds, high endurance, and low operating voltages, making it promising for neuromorphic applications~\cite{li2020red}. \item \textbf{MRAM}: Stores data using magnetic states rather than electric charges. MRAM provides fast read and write speeds, non-volatility, and unlimited endurance, suitable for applications requiring frequent weight updates~\cite{hu2019spin}. \end{itemize}

\subsubsection{High-Bandwidth Memory (HBM)}

HBM is a high-performance RAM interface for 3D-stacked DRAM integrated with processors~\cite{pedroni2019memory}. HBM offers significantly higher bandwidth compared to traditional DRAM due to its wide interface and close physical proximity to the processor. It also consumes less power per bit transferred, making it suitable for data-intensive neuromorphic applications.

\subsection{Hybrid Memory Systems}

Combining different memory technologies to form hybrid memory systems can leverage the advantages of each type~\cite{li2017utility, sun2023full, chi2024hybrid}. For instance, using SRAM for critical on-chip storage and HBM or DRAM for large off-chip storage can balance speed and capacity requirements. Non-volatile memories can be integrated for data persistence and in-memory computation capabilities.

\begin{table}[h]
\centering
\caption{Summary of Memory Technologies for Neuromorphic Computing}
\label{tab:memory_summary}
\begin{tabular}{lcccc}
\hline
\textbf{Memory Type} & \textbf{Latency} & \textbf{Bandwidth} & \textbf{Power} & \textbf{Density} \\
\hline
SRAM   & Very Low  & High       & High     & Low      \\
DRAM   & Moderate  & Moderate   & Moderate & High     \\
Flash  & High      & Low        & Low      & Very High \\
PCM    & Moderate  & Moderate   & Moderate & High     \\
ReRAM  & Low       & High       & Low      & High     \\
MRAM   & Low       & High       & Low      & Moderate \\
HBM    & Low       & Very High  & Low      & High     \\
\hline
\end{tabular}
\label{table2}
\end{table}

Table~\ref{table2} summarizes the key characteristics of the discussed memory technologies. The selection of memory type depends on the specific requirements of the neuromorphic application, such as the need for speed, capacity, energy efficiency, and the nature of data access patterns. Pedroni et al.~\cite{hbm_neuromorphic} implemented a high-performance SNN on an FPGA using HBM to store synaptic weights. By leveraging the high bandwidth of HBM, they achieved significant speedups in data transfer rates, reducing latency in weight-loading operations. The use of HBM, combined with pipelining techniques, resulted in enhanced throughput and energy efficiency compared to traditional DRAM-based implementations. Yang et al.~\cite{coreset} proposed a memory hierarchy system incorporating HBM and SRAM for FPGA-based CNN accelerators. HBM was used to store large filter weights and feature maps, while SRAM served as on-chip caches for frequently accessed data. This approach alleviated fan-in/fan-out constraints and improved data reuse, leading to higher processing speeds and reduced power consumption. Comparing these devices emphasizes the trade-offs between flexibility, performance, and energy efficiency in neuromorphic computing. It is essential to characterize flash memory error mechanisms in detail in order to develop models that improve error tolerance in sub-20nm flash memories. There is significant potential for application- and data-aware management mechanisms to enhance flash memory performance, similar to techniques used for DRAM and emerging memory technologies. These strategies will support the effective scaling of flash memory in the future, ensuring continued progress in neuromorphic computing on FPGA platforms.

\begin{figure}
    \graphicspath{ {D:\Stack} }
    \center \includegraphics[width=0.50\textwidth]{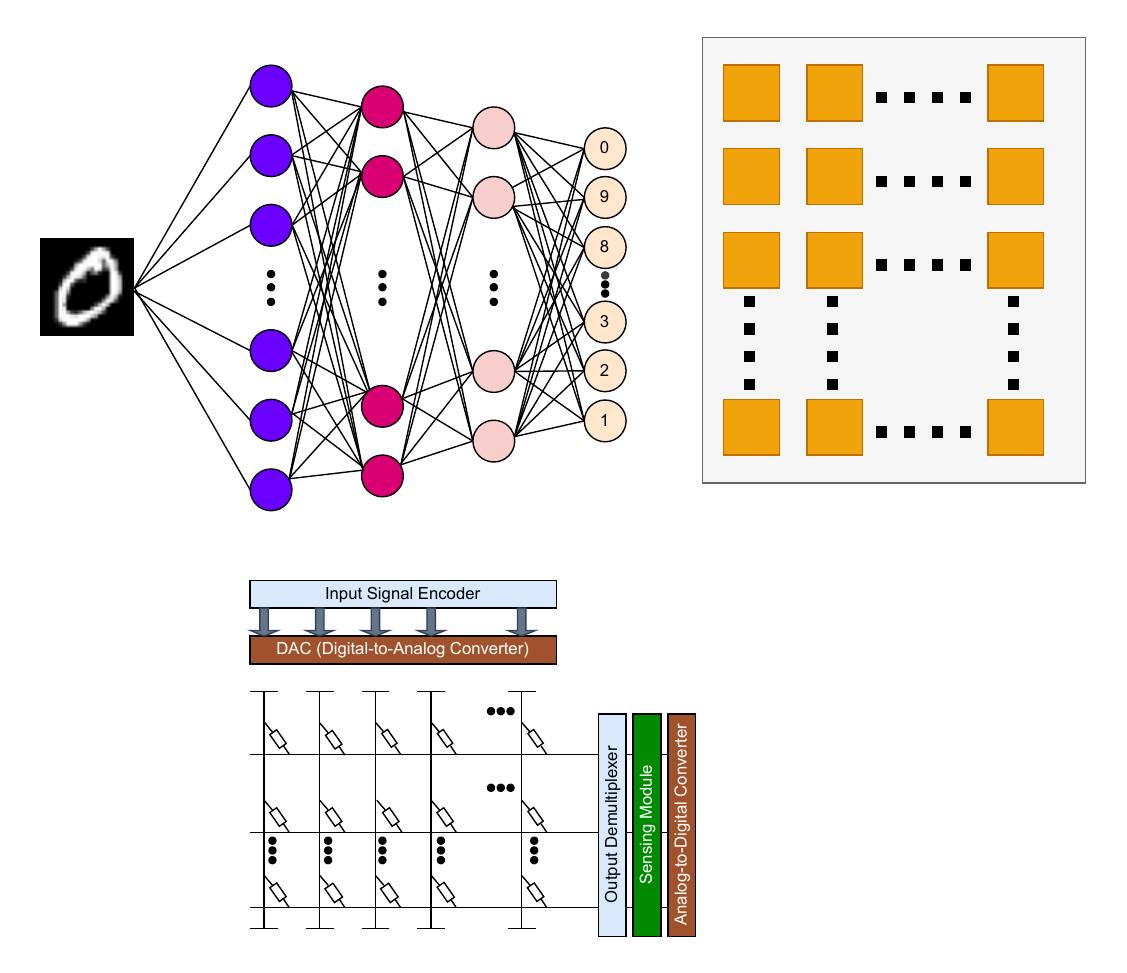}
    \caption{Neuromorphic Crossbar Architecture with Signal Encoding and Conversion Modules.  }
    \label{test}
\end{figure}

Fig. \ref{test} illustrates a neuromorphic computing crossbar architecture, the core structure for neural network implementations. The system includes signal encoding and conversion modules to enable efficient data transfer between digital and analog domains. This design highlights key components for processing in both domains, facilitating faster and more energy-efficient simulation of neural networks compared to digital-only systems. A notable case study involves implementing SNNs on FPGA platforms using SRAM for on-chip storage. The high speed and low latency of SRAM make it ideal for storing synaptic weights and neuronal states, ensuring rapid access to real-time spiking behavior.

\section{Challenges and Insights from Case Studies}

Neuromorphic computing, inspired by the brain’s architecture, offers significant performance and efficiency improvements over traditional methods. Memory technologies in FPGA-based neuromorphic systems face challenges similar to those in ASICs. Neuromorphic computing introduces analog paradigms, requiring changes to the hardware-software stack, including programming languages, asynchronous operations, and compiler optimizations. Researchers highlight the importance of scaling memory technologies while maintaining performance as system complexity grows \cite{soldavini2021survey}. Speed, bandwidth, and latency are critical factors in memory architecture. On-chip SRAM provides fast access but at higher costs and lower densities, while DRAM offers higher capacity with increased latency and power consumption \cite{Soldavini_2021}. Power optimization remains essential, as memory systems significantly affect energy consumption and heat dissipation. Emerging non-volatile memories, like ReRAM and PCM, provide more energy-efficient alternatives \cite{zahoor2020resistive} \cite{MOHSENI2023103008}. Reliability is also crucial. SRAM is more reliable than DRAM due to its lack of refresh cycles, but DRAM provides better density. PCM, with its ability to mimic biological neurons, and ReRAM, supporting stochastic behavior, offer integration potential for neuromorphic architectures \cite{upadhyay2019}. 

\subsection{Key Insights from Case Studies}

By addressing these challenges, FPGA-based neuromorphic computing systems can optimize memory architectures, advancing performance, scalability, and energy efficiency:

\begin{itemize}
    \item \textbf{Speed vs. Capacity}: SRAM offers low latency but limited capacity, while DRAM provides higher density with latency trade-offs.
    \item \textbf{Emerging Memories Integration}: Non-volatile memories require careful handling of compatibility and variability issues.
    \item \textbf{Memory Hierarchies}: Combining SRAM, DRAM, and emerging memories balances performance and efficiency.
    \item \textbf{Adaptive Memory Management}: Intelligent controllers optimize memory usage, enhancing system performance.
    \item \textbf{Application-Specific Alignment}: Memory selection should match the specific needs of the application, focusing on real-time processing, learning capabilities, and energy constraints.
\end{itemize}

\section{Conclusions}
This paper has provided a comprehensive review of memory technologies employed in FPGA-based neuromorphic computing systems. We discussed various research topics and proposed strategies to enhance FPGA memory scalability at both the system and architectural levels. Our analysis of trade-offs between different memory types—such as SRAM, DRAM, HBM, and emerging non-volatile memories—highlights essential considerations for implementing bio-inspired architectures effectively. This review shows researchers practical guidance on design initiation, expectations at the top system layer, and identifying hardware-level requirements for neuromorphic computing. We addressed unresolved research challenges related to reconfigurable circuits, outlining both current accomplishments and areas for further exploration. We also examined the impact of memory performance on neural network outputs, underscoring the importance of selecting suitable memory architectures to overcome the memory wall and enhance system efficiency and scalability.

\section{Trends and Future Predictions}
Neuromorphic computing on FPGA platforms is advancing rapidly, driven by improvements in memory technologies and architectures. A key trend is the integration of non-volatile memories like ReRAM, PCM, and MRAM, offering high density, low latency, and in-memory computing. These technologies reduce energy use by eliminating the need for continuous power, making them ideal for energy-efficient systems. However, challenges remain in integrating them with FPGA infrastructure and ensuring compatibility with neuromorphic processors. In-memory computing will play a critical role by minimizing data movement and improving energy efficiency. Advancements in 3D memory technologies, such as HBM and HMC, will further increase memory bandwidth and capacity, with future efforts focusing on thermal management and interconnect optimization. Energy-efficient memory solutions will remain a priority, with innovations like ultra-low-power memory cells and asynchronous operations to minimize power consumption. Hybrid memory systems—combining SRAM, DRAM, and non-volatile memory—will gain traction, optimized by intelligent memory controllers for dynamic data management. New programming tools and models, including high-level abstractions and domain-specific languages, will simplify the deployment of neural networks on FPGA platforms. Scalability will be addressed through modular design, using repeatable memory and processing units to manage system complexity. Integration with mainstream machine learning frameworks will accelerate adoption by bridging neuromorphic platforms with existing tools.

\bibliographystyle{IEEEtran}
\bibliography{external}

\end{document}